\documentclass[12pt,oneside,english,thmsa]{article}
\usepackage{amssymb}
\usepackage{amsmath}
\usepackage[T1]{fontenc}
\usepackage[latin1]{inputenc}
\makeatletter
 \newtheorem{thm}{Theorem}%[section]
 \newtheorem{rem}[thm]{Remark}
 \newtheorem{prop}[thm]{Proposition} %%Delete [thm] to re-start numbering

\newtheorem{theorem}{Theorem}%[section]

\newtheorem{definition}[theorem]{Definition}

\newtheorem{remark}[theorem]{Remark}

\makeatother
\begin{document}
\author{\textbf{Margarida de Faria}\\
CCM, University of Madeira, P-9000-390 Funchal\\
mfaria@uma.pt \and 
\textbf{Maria Jo\~{a}o Oliveira}\\
Univ.~Aberta, P-1269-001 Lisbon; \\
GFMUL, University of Lisbon, P-1649-003 Lisbon;\\
BiBoS, University of Bielefeld, D-33501 Bielefeld\\
oliveira@cii.fc.ul.pt \and \textbf{Ludwig Streit}\\
BiBoS, University of Bielefeld, D-33501 Bielefeld;\\
CCM, University of Madeira, P-9000-390 Funchal\\
streit@physik.uni-bielefeld.de}
\title{Feynman Integrals for Non-Smooth and Rapidly Growing Potentials}
\date{}
\maketitle

\begin{abstract}
The Feynman integral for the Schr\"odinger propagator is constructed
as a generalized function of white noise, for a linear space of potentials
spanned by measures and Laplace transforms of measures, i.e. locally
singular as well as rapidly growing at infinity. Remarkably, all these 
propagators admit a perturbation expansion.
\end{abstract}
\maketitle

\newpage

\section*{I. Introduction}

On a mathematical level of rigor, the construction of Feynman integrals
for quantum mechanical propagators will have to be done for specific
classes of potentials. In particular, the Feynman integrand has been
identified as a well-defined generalized function in white noise space,
e.g. for the following classes of potentials:

- (signed) finite measures which are ''small'' at infinity \cite{KS92,LLSW93}

- Fourier transforms of measures \cite{W95}

- Laplace transforms of finite measures \cite{KSW98}.

Potentials in the third space are locally smooth but may grow rapidly
at infinity, a prominent example is the Morse potential. On the other
hand the first of these classes includes locally singular potentials
such as the Dirac delta function. It is also important for the construction
of Feynman integrals with boundary conditions \cite{bbs}. Hence it
would be desirable to admit potentials which are linear combinations
of elements from the first and third space. The present paper addresses
this problem: we show the existence of Feynman integrals solving the 
propagator equation for such potentials.

\section*{II. White noise analysis}

In this section we briefly recall the concepts and results of white
noise analysis used throughout this work (see, e.g., \cite{BeKo88},
\cite{Hid75}, \cite{HKPS93}, \cite{Ko75}, \cite{KuTa80a}, \cite{KuTa80},
\cite{Kuo96}, \cite{Ob94} for a detailed explanation). 

The starting point of (one-dimensional) white noise analysis is the
real Gelfand triple \[
S(\Bbb{R})\subset L^{2}(\Bbb{R})\subset S'(\Bbb{R}),\]
 where $L^{2}:=L^{2}(\Bbb{R})$ is the real Hilbert space of all square
integrable functions w.r.t. the Lebesgue measure, $\mathcal{S}:=S(\Bbb{R})$
and $\mathcal{S}':=S'(\Bbb{R})$ are the real Schwartz spaces of test
functions and tempered distributions, respectively. In the sequel
we denote the norm on $L^{2}$ by $|\cdot|$, the corresponding inner
product by $(\cdot,\cdot)$, and the dual pairing between $\mathcal{S}'$
and $\mathcal{S}$ by $\left\langle \cdot,\cdot\right\rangle $. The
dual pairing $\left\langle \cdot,\cdot\right\rangle $ and the inner
product $(\cdot,\cdot)$ are connected by \[
\left\langle f,\xi\right\rangle =(f,\xi),\quad f\in L^{2},\xi\in\mathcal{S}.\]
 By $\{|\cdot|_{p}\}_{p\in\Bbb{N}}$ we denote a family of Hilbert
norms topologizing the space $\mathcal{S}$. 

Let $\mathcal{B}$ be the $\sigma$-algebra generated by the cylinder
sets on $\mathcal{S}'$. Through the Minlos theorem one may define
the white noise measure space $(\mathcal{S}',\mathcal{B},\mu)$ by
giving the characteristic function \[
C(\xi):=\int_{\mathcal{S}'}e^{i\left\langle \omega,\xi\right\rangle }\, d\mu(\omega)=e^{-\frac{1}{2}|\xi|^{2}},\quad\xi\in\mathcal{S}.\]
 Within this formalism a version of the (one-dimensional) Wiener Brownian
motion is given by \[
B(t):=\left\langle \omega,1\!\!1_{[0,t)}\right\rangle ,\quad\omega\in\mathcal{S}',\]
 where $1\!\!1_{A}$ denotes the indicator function of a set $A$. 

Now let us consider the complex Hilbert space $L^{2}(\mu):=L^{2}(\mathcal{S}',\mathcal{B},\mu)$.
As this space quite often shows to be too small for applications,
to proceed further we shall construct a Gelfand triple around the
space $L^{2}(\mu)$. More precisely, first we shall choose a space
of white noise test functions contained in $L^{2}(\mu)$ and then
we work on its larger dual space of distributions. In our case we
will use the space $\left(\mathcal{S}\right)^{-1}$ of generalized
white noise functionals or Kondratiev distributions and its well-known
subspace $\left(\mathcal{S}\right)'$ of Hida distributions (or generalized
Brownian functionals) with corresponding Gelfand triples \[
\left(\mathcal{S}\right)^{1}\subset L^{2}(\mu)\subset\left(\mathcal{S}\right)^{-1}\]
 and \[
\left(\mathcal{S}\right)\subset L^{2}(\mu)\subset\left(\mathcal{S}\right)'.\]
 Instead of reproducing the explicit construction of $\left(\mathcal{S}\right)^{-1}$
and $\left(\mathcal{S}\right)'$ (see, e.g., \cite{BeKo88}, \cite{HKPS93}),
in Theorems \ref{Prop6} and \ref{Prop3} below we will define both
spaces by their $T$-transforms. Given a $\Phi\in\left(\mathcal{S}\right)^{-1}$,
there exist $p,q\in\Bbb{N}_{0}$ such that we can define for every
\[
\xi\in U_{p,q}:=\{\xi\in\mathcal{S}:2^{q}\left|\xi\right|_{p}^{2}<1\}\]
 the $T$-transform of $\Phi$ by \begin{equation}
T\Phi(\xi):=\left\langle \!\left\langle \Phi,\exp(i\left\langle \cdot,\xi\right\rangle )\right\rangle \!\right\rangle .\label{1.7}\end{equation}
 Here $\left\langle \!\left\langle \cdot,\cdot\right\rangle \!\right\rangle $
denotes the dual pairing between $\left(\mathcal{S}\right)^{-1}$
and $\left(\mathcal{S}\right)^{1}$ which is defined as the bilinear
extension of the inner product on $L^{2}(\mu)$. In particular, for
Hida distributions $\Phi$, definition (\ref{1.7}) extends to $\xi\in\mathcal{S}$.
By analytic continuation, the definition of $T$-transform may be
extended to the underlying complexified space $\mathcal{S}_{\Bbb{C}}$
of $\mathcal{S}$. 

In order to define the spaces $\left(\mathcal{S}\right)^{-1}$ and
$\left(\mathcal{S}\right)'$ through their $T$-transforms we need
the following two definitions. 

\begin{definition}
\label{Def2}A function \(F:U\rightarrow \Bbb{C}\) is holomorphic on an open set
\(U\subset \mathcal{S}_{\Bbb{C}}\) if\newline
1. for all \(\theta_0\in U\) and any \(\theta\in \mathcal{S}_{\Bbb{C}}\) the 
mapping \(\Bbb{C}\ni \lambda\longmapsto F(\lambda \theta + \theta_0)\) is 
holomorphic on some neighborhood of \(0\in\Bbb{C}\),\newline 
2. \(F\) is locally bounded.
\end{definition}

\begin{definition}
\label{Def1}A function \(F:\mathcal{S}\rightarrow \Bbb{C}\) is called a 
\(U\)-functional whenever\newline
1. for every \(\xi_1,\xi_2\in \mathcal{S}\) the mapping 
\(\Bbb{R\ni \lambda }\longmapsto F(\lambda \xi_1+\xi_2)\) has an entire 
ex\-ten\-sion to \(\lambda \in \Bbb{C}\),\newline
2. there exist constants \(K_1,K_2>0\) such that 
\[
\left| F(z\xi)\right| \leq K_1\exp \left( K_2\left| z\right| ^2\left\|
\xi\right\| ^2\right) ,\quad \forall \,z\in \Bbb{C},\xi\in \mathcal{S}
\]
for some continuous norm \(\left\| \cdot \right\|\) on \(\mathcal{S}\).
\end{definition}

We are now ready to state the aforementioned characterization results. 

\begin{thm}
\label{Prop6}{\textrm{(\cite{KLS96})}} Let \(0\in U\subset \mathcal{S}_{\Bbb{C}}\) 
be an open set and \(F:U\rightarrow \Bbb{C}\) be a holomorphic function on \(U\). 
Then there is a unique \(\Phi\in \left(\mathcal{S}\right)^{-1}\) such that 
\(T\Phi = F\). 
Conversely, given a \(\Phi\in \left(\mathcal{S}\right)^{-1}\) the function 
\(T\Phi\) is holomorphic on some open set in \(\mathcal{S}_{\Bbb{C}}\) containing 
0. The correspondence between \(F\) and \(\Phi\) is a bijection if one identifies 
holomorphic functions which coincide on some open neighborhood of 0 in 
\(\mathcal{S}_{\Bbb{C}}\).
\end{thm}

\begin{thm}
\label{Prop3}{\textrm{(\cite{KLPSW96}, \cite{PS91})}}  The \(T\)-transform defines a 
bijection between the space \(\left(\mathcal{S}\right)'\) and the space of 
\(U\)-functionals.
\end{thm}
As a consequence of Theorem \ref{Prop6} one may derive the next two
statements. The first one concerns the convergence of sequences of
generalized white noise functionals and the second one the Bochner
integration of families of the same type of generalized functionals.
Similar results exist for Hida distributions (see, e.g., \cite{HKPS93}).

\begin{thm}
\label{Prop4}Let \(\left( \Phi _n\right) _{n\in \Bbb{N}}\) be a sequence in 
\(\left( \mathcal{S}\right) ^{-1}\) such that there are \(p,q\in \Bbb{N}_0\) so
that\newline
1. all \(T\Phi _n\) are holomorphic on 
\(U_{p,q}:=\{\theta \in \mathcal{S}_{\Bbb{C}}:2^q\left| \theta \right| _p^2<1\}\),\newline
2. there exists a \(C>0\) such that \(\left| T\Phi _n(\theta )\right| \leq C\)
for all \(\theta \in U_{p,q}\) and all \(n\in \Bbb{N}\),\newline
3. \(\left( T\Phi _n(\theta )\right) _{n\in \Bbb{N}}\) is a Cauchy sequence in 
\(\Bbb{C}\) for all \(\theta \in U_{p,q}\).\newline
Then \(\left( \Phi _n\right) _{n\in \Bbb{N}}\) converges strongly in \(\left( 
\mathcal{S}\right) ^{-1}\).
\end{thm}

\begin{thm}
\label{Prop5}Let \((\Lambda ,\mathcal{F},\nu )\) be a measure space and \(\lambda \longmapsto \Phi_\lambda\) be a mapping from \(\Lambda\) to \(\left( \mathcal{S}\right) ^{-1}\). We assume that there exists a \(U_{p,q}\subset \mathcal{S}_{\Bbb{C}}\), \(p,q\in \Bbb{N}_0\), such that\newline
1. \(T\Phi_\lambda\) is holomorphic on \(U_{p,q}\) for every \(\lambda \in \Lambda \),\newline
2. the mapping \(\lambda \longmapsto T\Phi_\lambda(\theta )\) is
measurable for every \(\theta \in U_{p,q}\),\newline
3. there is a \(C\in L^1(\Lambda ,\mathcal{F},\nu )\) such that \[ \left| T\Phi_\lambda(\theta )\right| \leq C(\lambda ),\quad \forall
\,\theta \in U_{p,q},\,\nu -\mathit{a.a.\/}\, \lambda \in \Lambda . \] Then there exist \(p^{\prime },q^{\prime }\in \Bbb{N}_0\), which only depend on \(p,q\), such that \(\Phi_\lambda\) is Bochner integrable. In particular, \[ \int_\Lambda \Phi_\lambda\,d\nu (\lambda )\in \left( \mathcal{S}\right)^{-1}
\] and \(T\left( \int_\Lambda \Phi_\lambda\,d\nu (\lambda )\right)\) is
holomorphic on \(U_{p^{\prime },q^{\prime }}\). One has \[ \left\langle\!\!\left\langle \int_\Lambda \Phi_\lambda\,d\nu (\lambda
),\varphi \right\rangle\!\!\right\rangle =\int_\Lambda \left\langle\!\left\langle \Phi_\lambda,\varphi \right\rangle\!\right\rangle\,d\nu
(\lambda ),\quad \forall \,\varphi \in \left( \mathcal{S}\right) ^1.
\]
\end{thm}

\section*{III. The free Feynman integral}

We follow \cite{FPS91} and \cite{HS83} in viewing the Feynman integral
as a weighted average over Brownian paths. We use a slight change
in the definition of the paths, which are here modeled by \[
x(\tau)=x-\sqrt{\frac{\hbar}{m}}\int_{\tau}^{t}\omega(s)\, ds:=x-\sqrt{\frac{\hbar}{m}}\left\langle \omega,1\!\!1_{(\tau,t]}\right\rangle ,\quad\omega\in\mathcal{S}'.\]
 That is, instead of fixing the starting point of the paths, we fix
the endpoint $x$ at time $t$. In the sequel we set $\hbar=m=1$.
Correspondingly, the Feynman integrand for the free motion is defined
by \[
I_{0}:=I_{0}(x,t|y,t_{0}):=N\exp\left(\frac{i+1}{2}\int_{\Bbb{R}}\omega^{2}(\tau)\, d\tau\right)\delta(x(t_{0})-y),\]
 where, informally, $N$ is a normalizing factor, more precisely,
$N\exp\left(\cdot\right)$ is a Gauss kernel (see, e.g., \cite{HKPS93},
\cite{LLSW93}). We recall that the Donsker delta function $\delta(x(t_{0})-y)$
is used to fix the starting point of the paths at time $t_{0}<t$.
The $T$-transform of the free Feynman integrand \begin{eqnarray}
TI_{0}(\xi) & = & \frac{1}{\sqrt{2\pi i(t-t_{0})}}\exp\left(\!-\frac{i}{2}\int_{\Bbb{R}}\xi^{2}(\tau)\, d\tau\right)\label{1.8}\\
 &  & \times\exp\left(\frac{i}{2(t-t_{0})}\left(\int_{t_{0}}^{t}\xi(\tau)\, d\tau+x-y\right)^{2}\right)\nonumber \end{eqnarray}
 is a $U$-functional and we use it to define $I_{0}$ as a Hida distribution
(see \cite{FPS91}). 

From the physical point of view, equality (\ref{1.8}) clearly shows
that the Feynman integral $TI_{0}(0)$ is the free particle propagator
\[
\frac{1}{\sqrt{2\pi i(t-t_{0})}}\exp\left(\frac{i}{2(t-t_{0})}(x-y)^{2}\right).\]
 Besides this particular case, even for nonzero $\xi$ the $T$-transform
of $I_{0}$ has a physical interpretation. Integrating formally by
parts we find \begin{eqnarray*}
TI_{0}(\xi) & = & \int_{\mathcal{S}'}I_{0}(\omega)\exp\left(-i\int_{t_{0}}^{t}x(\tau)\dot{\xi}(\tau)\, d\tau\right)d\mu(\omega)\\
 &  & \times\exp\left(-\frac{i}{2}\int_{[t_{0},t]^{c}}\xi^{2}(\tau)\, d\tau+ix\xi(t)-iy\xi(t_{0})\right).\end{eqnarray*}
 The term $\exp\left(-i\int_{t_{0}}^{t}x(\tau)\dot{\xi}(\tau)\, d\tau\right)$
would thus correspond to a time-dependent potential $W(x,t)=\dot{\xi}(t)x$.
In fact, it is straighforward to verify that \[
\Theta(t-t_{0})\cdot TI_{0}(\xi)=K_{0}^{(\xi)}\exp\left(-\frac{i}{2}\int_{[t_{0},t]^{c}}\xi^{2}(\tau)\, d\tau+ix\xi(t)-iy\xi(t_{0})\right),\]
 where $\Theta$ is the Heaviside function and \begin{eqnarray*}
K_{0}^{(\xi)}:=K_{0}^{(\xi)}(x,t|y,t_{0}) & := & \frac{\Theta(t-t_{0})}{\sqrt{2\pi i|t-t_{0}|}}\exp\left(-\frac{i}{2}\int_{t_{0}}^{t}\xi^{2}(\tau)\, d\tau\right)\\
 &  & \times\exp\left(\frac{i}{2|t-t_{0}|}\left(\int_{t_{0}}^{t}\xi(\tau)\, d\tau+x-y\right)^{2}\right)\\
 &  & \times\exp\left(iy\xi(t_{0})-ix\xi(t)\right)\end{eqnarray*}
 is the Green function corresponding to the potential $W$, i.e.,
$K_{0}^{(\xi)}$ obeys the Schr\"{o}dinger equation \begin{equation}
\left(i\partial_{t}+\frac{1}{2}\partial_{x}^{2}-\dot{\xi}(t)x\right)K_{0}^{(\xi)}(x,t|y,t_{0})=i\delta(t-t_{0})\delta(x-y).\label{free}\end{equation}

\section*{IV. Interactions}

In the sequel $\mathcal{K}_{1}$ denotes the linear space of all potentials
$V$ on $\Bbb{R}$ of the form \[
V(x)=\int_{\Bbb{R}}e^{\alpha x}\, dm(\alpha),\qquad x\in\Bbb{R},\]
 where $m$ is a complex measure on the Borel sets on $\Bbb{R}$ fulfilling
the condition \begin{equation}
\int_{\Bbb{R}}e^{C\left|\alpha\right|}\, d\left|m\right|(\alpha)<\infty,\qquad\forall\, C>0\label{1.4}\end{equation}
 (cf.~\cite{KSW98}), and $\mathcal{K}_{2}$ denotes the space of
all potentials $V$ on $\Bbb{R}$ which are generalized functions
of the type \[
V(x)=\int_{\Bbb{R}}\delta(x-y)\, dm(y),\qquad x\in\Bbb{R},\]
 where $dm(y):=V(y)dy$ is a finite signed Borel measure of bounded
support (cf.~\cite{KS92}). 

\begin{rem}
A Lebesgue dominated convergence argument shows that potentials in
$\mathcal{K}_{1}$ are restrictions to the real line of entire functions
\cite{KSW98}. In particular, they are locally bounded and smooth.
\end{rem}
Our aim is to define the Feynman integrand \begin{equation}
I:=I_{0}\cdot\exp\left(-i\int_{t_{0}}^{t}V(x(\tau))\, d\tau\right)\label{1.1}\end{equation}
 for a potential $V$ of the form $V=V_{1}+V_{2}$, $V_{i}\in\mathcal{K}_{i}$,
\begin{equation}
V_{1}(x)=\int_{\Bbb{R}}e^{\alpha x}\, dm_{1}(\alpha),\quad V_{2}(x)=\int_{\Bbb{R}}\delta(x-y)\, dm_{2}(y),\label{1.5}\end{equation}
 where \[
x(\tau)=x-\int_{\tau}^{t}\omega(s)\, ds,\quad\omega\in\mathcal{S}',\]
 as before. In order to do this, first we must give a meaning to the
heuristic expression (\ref{1.1}). In Theorem \ref{Prop2} it will
be shown that $I$ is indeed a well-defined generalized white noise
functional. Secondly, it has to be proven that the expectation of
$I$ solves the Schr\"{o}dinger equation for the potential $V$. 

As a first step we expand the exponential in (\ref{1.1}) into a perturbation
series. This leads to \begin{eqnarray}
 &  & I=\sum_{n=0}^{\infty}\frac{(-i)^{n}}{n!}\sum_{k=0}^{n}\binom{n}{k}k!\int_{\Delta_{k}}d^{k}\tau\int_{t_{0}}^{t}d^{n-k}s\nonumber \\
 &  & \int_{\Bbb{R}^{k}}\int_{\Bbb{R}^{n-k}}I_{0}\exp\left(\sum_{l=1}^{n-k}\alpha_{l}x(s_{l})\right)\prod_{j=1}^{k}\delta(x(\tau_{j})-x_{j})\prod_{l=1}^{n-k}dm_{1}(\alpha_{l})\prod_{j=1}^{k}dm_{2}(x_{j}),\nonumber \\
\label{1.10}\end{eqnarray}
where $\Delta_{k}:=\{(\tau_{1},...,\tau_{k}):t_{0}<\tau_{1}<...<\tau_{k}<t\}$.
In the above expression the integrals over $\Delta_{k},\Bbb{R}^{k}$
and $\left[t_{0},t\right]^{n-k},\Bbb{R}^{n-k}$ disappear, respectively,
for $k=0$ and $k=n$. Our aim is to apply Theorems \ref{Prop4} and
\ref{Prop5} to show the existence of the above series and integrals.
However, first we have to establish the pointwise multiplication of
generalized functionals \[
I_{0}\exp\left(\sum_{l=1}^{n-k}\alpha_{l}x(s_{l})\right)\prod_{j=1}^{k}\delta(x(\tau_{j})-x_{j})\]
 as a well-defined generalized functional. Due to the characterization
result Theorem \ref{Prop3} it is enough to define this product through
its $T$-transform. Arguing informally, for $\xi\in\mathcal{S}$ we
are led to \begin{eqnarray*}
 &  & T\left(I_{0}\exp\left(\sum_{l=1}^{n-k}\alpha_{l}x(s_{l})\right)\prod_{j=1}^{k}\delta(x(\tau_{j})-x_{j})\right)(\xi)\\
 & = & \int_{\mathcal{S}^{\prime}}I_{0}\exp\left(\sum_{l=1}^{n-k}\alpha_{l}x(s_{l})\right)\prod_{j=1}^{k}\delta(x(\tau_{j})-x_{j})\exp\left(i\left\langle \omega,\xi\right\rangle \right)d\mu(\omega)\\
 & = & \exp\left(x\sum_{l=1}^{n-k}\alpha_{l}\right)\cdot T\left(I_{0}\prod_{j=1}^{k}\delta(x(\tau_{j})-x_{j})\right)(\xi+i\sum_{l=1}^{n-k}\alpha_{l}1\!\!1_{(s_{l},t]}).\end{eqnarray*}
 The product $I_{0}\prod_{j=1}^{k}\delta(x(\tau_{j})-x_{j})$ is a
slight generalization of the free Feynman integrand $I_{0}$, with
more than just one delta function, and may be defined by its $T$-transform,
\begin{eqnarray}
 &  & T\left(I_{0}\prod_{j=1}^{k}\delta(x(\tau_{j})-x_{j})\right)(\xi)\nonumber \\
 & = & \exp\left(-\frac{i}{2}\int_{[t_{0},t]^{c}}\xi^{2}(s)ds+ix\xi(t)-iy\xi(t_{0})\right)\prod_{j=1}^{k+1}K_{0}^{(\xi)}(x_{j},\tau_{j}|x_{j-1},\tau_{j-1})\nonumber \\
 & = & \exp\left(-\frac{i}{2}\int_{\Bbb{R}}\xi^{2}(s)ds\right)\prod_{j=1}^{k+1}\left\{ \frac{1}{\sqrt{2\pi i(\tau_{j}-\tau_{j-1})}}\right.\nonumber \\
 &  & \left.\times\exp\left(\frac{i}{2\left(\tau_{j}-\tau_{j-1}\right)}\left(\int_{\tau_{j-1}}^{\tau_{j}}\xi(s)ds+x_{j}-x_{j-1}\right)^{2}\right)\right\} .\nonumber \\
\label{1.2}\end{eqnarray}
 Here $\tau_{0}:=t_{0},x_{0}:=y,\tau_{k+1}:=t$, and $x_{k+1}:=x$.
Clearly the explicit formula (\ref{1.2}) is continuously extendable
to all $\xi\in L^{2}$ which allows an extension of $T\left(I_{0}\prod_{j=1}^{k}\delta(x(\tau_{j})-x_{j})\right)$
to the argument $\xi+i\sum_{l=1}^{n-k}\alpha_{l}1\!\!1_{(s_{l},t]}$. 

\begin{prop}
\label{Prop1}The product \[ \Phi _{n,k}:=I_0\exp \left( \sum_{l=1}^{n-k}\alpha _lx(s_l)\right)
\prod_{j=1}^k\delta (x(\tau _j)-x_j) 
\]
defined by 
\begin{eqnarray*}
&&T\Phi _{n,k}(\xi ) \\
&=&\!\!T\left( I_0\prod_{j=1}^k\delta (x(\tau _j)-x_j)\right) \left( \xi
+i\sum_{l=1}^{n-k}\alpha _l1\!\!1_{(s_l,t]}\right) \exp\!\! \left(
x\sum_{l=1}^{n-k}\alpha _l\right) \\
&=&\!\!\exp\!\! \left( -\frac i2\int_{\Bbb{R}}\left( \xi
(s)+i\sum_{l=1}^{n-k}\alpha _l1\!\!1_{(s_l, t]}(s)\right) ^2ds\right)
\prod_{j=1}^{k+1}\frac 1{\sqrt{2\pi i(\tau _j-\tau _{j-1})}}\\
&&\!\!\!\!\times \exp\!\! \left( \!\sum_{j=1}^{k+1}\frac i{2\left( \tau _j-\tau
_{j-1}\right) }\!\left( \!\int_{\tau _{j-1}}^{\tau _j}\!\!\left( \xi
(s)+i\sum_{l=1}^{n-k}\alpha _l1\!\!1_{(s_l,t]}(s)\!\!\right)
\!ds+x_j-x_{j-1}\!\right) ^2\right) \\
&&\!\!\!\!\times \exp\!\! \left( x\sum_{l=1}^{n-k}\alpha _l\right)
\end{eqnarray*}
is a Hida distribution.
\end{prop}
\noindent \textbf{Proof.} It is obvious that the latter explicit formula
fulfills the first part of Definition \ref{Def1}, analyticity. In
order to prove that $\Phi_{n,k}$ is a Hida distribution by application
of Theorem \ref{Prop3}, we only have to show that $T\Phi_{n,k}$
also obeys a bound as in the second part of Definition \ref{Def1}.
For every $\theta\in\mathcal{S}_{\Bbb{C}}$ we have \begin{eqnarray*}
 &  & \left|T\Phi_{n,k}(\theta)\right|\\
 & \leq & \exp\!\!\left(\left|x\right|\sum_{l=1}^{n-k}\left|\alpha_{l}\right|\right)\\
 &  & \times\prod_{j=1}^{k+1}\frac{1}{\sqrt{2\pi(\tau_{j}-\tau_{j-1})}}\left|\exp\!\!\left(-\frac{i}{2}\int_{\Bbb{R}}\theta^{2}(s)ds+\sum_{l=1}^{n-k}\alpha_{l}\int_{\Bbb{R}}\theta(s)1\!\!1_{(s_{l},t]}(s)ds\right)\right|\\
 &  & \times\left|\exp\!\!\left(\sum_{j=1}^{k+1}\frac{i}{2\left(\tau_{j}-\tau_{j-1}\right)}\left(\int_{\tau_{j-1}}^{\tau_{j}}\theta(s)ds\right)^{2}\right)\right|\\
 &  & \times\left|\exp\!\!\left(\sum_{j=1}^{k+1}\frac{1}{\tau_{j-1}-\tau_{j}}\left(\int_{\tau_{j-1}}^{\tau_{j}}\theta(s)ds\right)\sum_{l=1}^{n-k}\alpha_{l}\left(\int_{\tau_{j-1}}^{\tau_{j}}1\!\!1_{(s_{l},t]}(s)ds\right)\right)\right|\\
 &  & \times\left|\exp\!\!\left(\sum_{j=1}^{k+1}\frac{i\left(x_{j}-x_{j-1}\right)}{\tau_{j}-\tau_{j-1}}\int_{\tau_{j-1}}^{\tau_{j}}\left(\theta(s)+i\sum_{l=1}^{n-k}\alpha_{l}1\!\!1_{(s_{l},t]}(s)\right)ds\right)\right|\end{eqnarray*}
 which is majorized by 

\begin{eqnarray}
\left|T\Phi_{n,k}(\theta)\right| & \leq & \prod_{j=1}^{k+1}\frac{1}{\sqrt{2\pi(\tau_{j}-\tau_{j-1})}}\exp\left(2\left\Vert \theta\right\Vert ^{2}\right)\nonumber \\
 &  & \times\exp\left(\left(\left|x\right|+t-t_{0}+\left\Vert \theta\right\Vert ^{2}\right)\sum_{l=1}^{n-k}\left|\alpha_{l}\right|\right)\label{po}\\
 &  & \times\exp\left(4\max_{0\leq j\leq k+1}\left|x_{j}\right|\sum_{l=1}^{n-k}\left|\alpha_{l}\right|\right)\exp\left(\max_{0\leq j\leq k+1}\left(\left|x_{j}\right|^{2}\right)\right)\nonumber \\
 & =: & C(\tau_{1},...,\tau_{k};\alpha_{1},...,\alpha_{n-k};x_{1},...,x_{k};\theta)=:C\nonumber \end{eqnarray}

\noindent independent of $s_{1},...,s_{n-k}$, where \[
\left\Vert \theta\right\Vert :=\sup_{s\in\left[t_{0},t\right]}\left|\theta(s)\right|+\int_{t_{0}}^{t}\left|\dot{\theta}(s)\right|ds+\left|\theta\right|\]

\noindent is a continuous norm on $\mathcal{S}_{\Bbb{C}}$, cf.~Appendix
below. This estimate for $T\Phi_{n,k}$ is of the form required in
Definition 2, which completes the proof.\hfill{}$\blacksquare\medskip$

According to Proposition \ref{Prop1}, all $\Phi_{n,k}$ are Hida
distributions and thus also generalized white noise functionals with
$T\Phi_{n,k}$ entire on $\mathcal{S}_{\Bbb{C}}$. Moreover, each
$T\Phi_{n,k}(\theta)$ is a measurable function of $\tau_{1},...,\tau_{k};s_{1},...,s_{n-k};$
$\alpha_{1},...,\alpha_{n-k};x_{1},...,x_{k}$ for every $\theta\in\mathcal{S}_{\Bbb{C}}$.
Hence, in order to apply Theorem \ref{Prop5} to prove the existence
of the integrals in $I$, we only have to find a suitable integrable
bound for $\left|T\Phi_{n,k}(\theta)\right|$. Since the measure $m_{1}$
fulfills the integrability condition (\ref{1.4}) and the signed measure
$m_{2}$ is finite and has support contained in some bounded interval
$\left[-a,a\right]$, $a>0$, one may infer the integrability of $C$
for every $\theta\in\mathcal{S}_{\Bbb{C}}$: \begin{eqnarray*}
 &  & \left|\int_{\Delta_{k}}d^{k}\tau\int_{t_{0}}^{t}d^{n-k}s\int_{\Bbb{R}^{k}}\prod_{j=1}^{k}dm_{2}(x_{j})\int_{\Bbb{R}^{n-k}}\prod_{l=1}^{n-k}d\left|m_{1}\right|(\alpha_{l})C\right|\\
 & \leq & \exp\left(2\left\Vert \theta\right\Vert ^{2}+b^{2}\right)(t-t_{0})^{n-k}\\
 &  & \times\int_{\Delta_{k}}\prod_{j=1}^{k+1}\frac{1}{\sqrt{2\pi(\tau_{j}-\tau_{j-1})}}d^{k}\tau\left|\int_{\Bbb{R}}dm_{2}(x)\right|^{k}\\
 &  & \times\left(\int_{\Bbb{R}}\exp\left(\left(\left|x\right|+4b+t-t_{0}+\left\Vert \theta\right\Vert ^{2}\right)\left|\alpha\right|\right)d\left|m_{1}\right|(\alpha)\right)^{n-k},\end{eqnarray*}
 where $b:=\max\{ a,|y|,|x|\}$. Thus, according to Theorem \ref{Prop5},
there exists an open set $U\subset\mathcal{S}_{\Bbb{C}}$ independent
of $n$ such that \[
I_{n,k}:=\int_{\Delta_{k}}d^{k}\tau\int_{t_{0}}^{t}d^{n-k}s\int_{\Bbb{R}^{k}}\int_{\Bbb{R}^{n-k}}\Phi_{n,k}\prod_{l=1}^{n-k}dm_{1}(\alpha_{l})\prod_{j=1}^{k}dm_{2}(x_{j})\in\left(\mathcal{S}\right)^{-1}\]
 for each $k\leq n$ and every $n\in\Bbb{N}$, and all $TI_{n,k}$
are holomorphic on $U$. To conclude the existence of $I$ we only
have to prove that the series in $n$ converges in $\left(\mathcal{S}\right)^{-1}$
in the strong sense. This follows from Theorem \ref{Prop4}. In fact,
due to (\ref{1.10}), for every $\theta\in U$ one has \[
\left|TI(\theta)\right|\leq\sum_{n=0}^{\infty}\frac{1}{n!}\sum_{k=0}^{n}\binom{n}{k}k!\left|TI_{n,k}(\theta)\right|\]
 where the right-hand side is upper bounded by the factor $\exp\left(2\left\Vert \theta\right\Vert ^{2}+b^{2}\right)$
times the Cauchy product of the convergent series \begin{eqnarray*}
 &  & \left(\sum_{n=0}^{\infty}\frac{1}{n!}\left((t-t_{0})\int_{\Bbb{R}}e^{\left(\left|x\right|+4b+t-t_{0}+\left\Vert \theta\right\Vert ^{2}\right)\left|\alpha\right|}d\left|m_{1}\right|(\alpha)\right)^{n}\right)\\
 &  & \times\left(\sum_{n=0}^{\infty}\left|\int_{\Bbb{R}}dm_{2}(x)\right|^{n}\int_{\Delta_{n}}\prod_{j=1}^{n+1}\frac{1}{\sqrt{2\pi(\tau_{j}-\tau_{j-1})}}d^{n}\tau\right)\\
 & = & \exp\left((t-t_{0})\int_{\Bbb{R}}e^{\left(\left|x\right|+4b+t-t_{0}+\left\Vert \theta\right\Vert ^{2}\right)\left|\alpha\right|}d\left|m_{1}\right|(\alpha)\right)\\
 &  & \times\sum_{n=0}^{\infty}\left|\int_{\Bbb{R}}dm_{2}(x)\right|^{n}\int_{\Delta_{n}}\prod_{j=1}^{n+1}\frac{1}{\sqrt{2\pi(\tau_{j}-\tau_{j-1})}}d^{n}\tau.\end{eqnarray*}
 We note that the latter series converges because \[
\int_{\Delta_{n}}\prod_{j=1}^{n+1}\frac{1}{\sqrt{2\pi(\tau_{j}-\tau_{j-1})}}d^{n}\tau=\left(\frac{\Gamma\left(1/2\right)}{\sqrt{2\pi}}\right)^{n+1}\frac{\left(t-t_{0}\right)^{(n-1)/2}}{\Gamma\left(\frac{n+1}{2}\right)}\]
is rapidly decreasing in $n$. 

In this way we have proved the following result. 

\begin{thm}
\label{Prop2} For every $V_{1}\in\mathcal{K}_{1}$ and $V_{2}\in\mathcal{K}_{2}$
 of the form (\ref{1.5}), the \\
\begin{eqnarray*}
&&I:=\sum_{n=0}^\infty \frac{(-i)^n}{n!}\sum_{k=0}^n\binom nkk!\int_{\Delta
_k}d^k\tau \int_{t_0}^td^{n-k}s \\
&&\int_{\Bbb{R}^k}\int_{\Bbb{R}^{n-k}}I_0\exp \left( \sum_{l=1}^{n-k}\alpha
_lx(s_l)\right) \prod_{j=1}^k\delta (x(\tau
_j)-x_j)\prod_{l=1}^{n-k}dm_1(\alpha _l)\prod_{j=1}^kdm_2(x_j),
\end{eqnarray*}
exists as a generalized white noise functional. The series converges
strongly in \((\mathcal{S})^{-1}\) and the integrals exist in the sense of
Bochner integrals. Therefore we may express the \(T\)-transform of \(I\) by 
\begin{eqnarray*}
&&TI(\theta )=\sum_{n=0}^\infty \frac{(-i)^n}{n!}\sum_{k=0}^n\binom
nkk!\int_{\Delta _k}d^k\tau \int_{t_0}^td^{n-k}s \\
&&\!\!\!\!\int_{\Bbb{R}^k}\!\int_{\Bbb{R}^{n-k}}\!\!\!\!T\!\left( \! I_0\exp\!\! \left(
\sum_{l=1}^{n-k}\alpha _lx(s_l)\!\right) \!\prod_{j=1}^k\delta (x(\tau
_j)-x_j)\!\right) \!(\theta )\!\prod_{l=1}^{n-k}\!dm_1(\alpha
_l)\prod_{j=1}^k\!dm_2(x_j)
\end{eqnarray*}
for every \(\theta \) in a neighborhood 
\(\left\{ \theta \in \mathcal{S}_{\Bbb{C}}:2^q\left| \theta \right| _p^2<1\right\} \) of zero, for some \(p,q\in \Bbb{N}_0\)
\end{thm}
According to Theorem \ref{Prop2}, $I$ is a well-defined generalized
white noise functional. In order to conclude that $I$ defines a Feynman
integrand it remains to show that the expectation $TI(0)$ of $I$
solves the Schr\"{o}dinger equation for a potential $V=V_{1}+V_{2},V_{i}\in\mathcal{K}_{i}$.
As in the free motion case we consider, more generally,\[
K^{(\theta)}(x,t|y,t_{0}):=\Theta(t-t_{0})TI(\theta)\exp\left(\frac{i}{2}\int_{[t_{0},t]^{c}}\theta^{2}(\tau)\, d\tau+iy\theta(t_{0})-ix\theta(t)\right).\]
Insertion of $TI(\theta)$ as given in Theorem \ref{Prop2}, with\[
T\left(I_{0}\exp\left(\sum_{l=1}^{n-k}\alpha_{l}x(s_{l})\right)\prod_{j=1}^{k}\delta(x(\tau_{j})-x_{j})\right)\]
as in Proposition 4, yields

\[
K^{(\theta)}(x,t|y,t_{0})=\sum_{n=0}^{\infty}K_{n}^{(\theta)}(x,t|y,t_{0}),\]
 with\begin{eqnarray}
K_{n}^{(\theta)}(x,t|y,t_{0})\!\! & := & \!\!\frac{(-i)^{n}}{n!}\int_{t_{0}}^{t}d^{n}s\int_{\Bbb{R}^{n}}\prod_{l=1}^{n}dm_{1}(\alpha_{l})K_{0}^{(\theta_{n})}(x,t|y,t_{0})\nonumber \\
 &  & \!\!+\sum_{k=1}^{n-1}\frac{(-i)^{n-k}}{(n-k)!}\int_{t_{0}}^{t}d^{n-k}s\int_{\Bbb{R}^{n-k}}\prod_{l=1}^{n-k}dm_{1}(\alpha_{l})G_{k}^{(\theta_{n-k})}(x,t|y,t_{0})\nonumber \\
 &  & \!\!+G_{n}^{(\theta)}(x,t|y,t_{0}),\nonumber \\
\label{Eq1}\end{eqnarray}
where we have set $\theta_{n-k}:=\theta_{n-k}(s_{1},...,s_{n-k},\alpha_{1},...,\alpha_{n-k}):=\theta+i\sum_{l=1}^{n-k}\alpha_{l}1\!\!1_{(s_{l},t]}$
for $k=0,...,n-1$, $\theta_{0}:=\theta$, and \[
G_{k}^{(\theta_{n-k})}(x,t|y,t_{0}):=(-i)^{k}\!\!\int_{\Delta_{k}}\!\! d^{k}\tau\!\!\int_{\Bbb{R}^{k}}\prod_{j=1}^{k}dm_{2}(x_{j})\prod_{j=1}^{k+1}K_{0}^{(\theta_{n-k})}(x_{j},\tau_{j}|x_{j-1},\tau_{j-1})\]
for $k=1,...,n,\, n>0.$

We expect $K^{(\theta)}$ to be the propagator corresponding to the
potential $W(x,t)=V(x)+\dot{\theta}(t)x$. 

\begin{thm}
$K^{(\theta)}(x,t\vert y,t_{0})$ is a Green function for the Schroedinger
equation \begin{equation}
\left(i\partial_{t}+\frac{1}{2}\partial_{x}^{2}-\dot{\theta}(t)x-V(x)\right)K^{(\theta)}(x,t\vert y,t_{0})=i\delta(t-t_{0})\delta(x-y).\label{Eq5}\end{equation}
In particular, $K(x,t\vert y,t_{0}):=TI(0)$ is a Feynman integral
solving \begin{equation}
i\partial_{t}K(x,t\vert y,t_{0})=\left(-\frac{1}{2}\partial_{x}^{2}+V(x)\right)K(x,t\vert y,t_{0}),\quad\hbox{for }\, t>t_{0}.\label{FOS}\end{equation}
\end{thm}

\begin{remark} $K$ corresponds to a unitary evolution whenever 
$H= -\frac{1}{2}\partial_{x}^{2}+V$ has a unique self-adjoint extension.
\end{remark}

\noindent
\textbf{Proof.}
\noindent Let us consider an interval $\left[T_{0},T\right]$ such
that $\left[t_{0},t\right]\subset\left[T_{0},T\right]$. Estimates
similar to those done in the proof of Proposition \ref{Prop1} show
that $K_{n}^{(\theta)}(\cdot,\cdot|y,t_{0})$ is locally integrable
on $\Bbb{R}\times\left[T_{0},T\right]$ with respect to $dm_{2}\times dt$
and the Lebesgue measure. Therefore, we may regard $K_{n}^{(\theta)}$
as a distribution on $\mathcal{D}(\Omega):=\mathcal{D}(\Bbb{R}\times\left[T_{0},T\right])$:
\[
\left\langle K_{n}^{(\theta)}(\cdot,\cdot|y,t_{0}),\varphi\right\rangle =\int_{\Bbb{R}}dx\int_{T_{0}}^{T}dtK_{n}^{(\theta)}(x,t|y,t_{0})\varphi(x,t),\quad\varphi\in\mathcal{D}(\Omega).\]
 And we may also define a distribution $V_{2}K_{n}^{(\theta)}$ by
setting \[
\left\langle V_{2}K_{n}^{(\theta)}(\cdot,\cdot|y,t_{0}),\varphi\right\rangle =\int_{\Bbb{R}}dm_{2}(x)\int_{T_{0}}^{T}dtK_{n}^{(\theta)}(x,t|y,t_{0})\varphi(x,t),\quad\varphi\in\mathcal{D}(\Omega).\]
To abbreviate we introduce the notation $\hat{L}:=i\partial_{t}+\frac{1}{2}\partial_{x}^{2}-\dot{\theta}(t)x$
and $\hat{L}^{*}$ for the dual operator. According to (\ref{Eq1}),
observe that for any test function $\varphi\in\mathcal{D}(\Omega)$
one finds \begin{eqnarray}
 &  & \left\langle \hat{L}K_{n}^{(\theta)},\varphi\right\rangle \nonumber \\
 & = & \!\!\!\frac{(-i)^{n}}{n!}\left\langle \int_{t_{0}}^{\cdot}d^{n}s\int_{\Bbb{R}^{n}}\prod_{l=1}^{n}dm_{1}(\alpha_{l})K_{0}^{(\theta_{n})}(\cdot,\cdot|y,t_{0}),\hat{L}^{*}\varphi\right\rangle \nonumber \\
 &  & \!\!\!+\sum_{k=1}^{n-1}\frac{(-i)^{n-k}}{(n-k)!}\left\langle \int_{t_{0}}^{\cdot}d^{n-k}s\int_{\Bbb{R}^{n-k}}\prod_{l=1}^{n-k}dm_{1}(\alpha_{l})G_{k}^{(\theta_{n-k})}(\cdot,\cdot|y,t_{0}),\hat{L}^{*}\varphi\right\rangle \label{Eq2}\\
 &  & \!\!\!+\left\langle G_{n}^{(\theta)}(\cdot,\cdot|y,t_{0}),\hat{L}^{*}\varphi\right\rangle ,\nonumber \end{eqnarray}
 where \begin{eqnarray}
 &  & \frac{(-i)^{n}}{n!}\left\langle \int_{t_{0}}^{\cdot}d^{n}s\int_{\Bbb{R}^{n}}\prod_{l=1}^{n}dm_{1}(\alpha_{l})K_{0}^{(\theta_{n})}(\cdot,\cdot|y,t_{0}),\hat{L}^{*}\varphi\right\rangle \label{Eq3}\\
 & = & \frac{(-i)^{n-1}}{(n-1)!}\left\langle V_{1}\int_{t_{0}}^{\cdot}d^{n-1}s\int_{\Bbb{R}^{n-1}}\prod_{l=1}^{n-1}dm_{1}(\alpha_{l})K_{0}^{(\theta_{n-1})}(\cdot,\cdot|y,t_{0}),\varphi\right\rangle \nonumber \end{eqnarray}
 cf.~\cite{KSW98}, and \begin{equation}
\left\langle G_{n}^{(\theta)}(\cdot,\cdot|y,t_{0}),\hat{L}^{*}\varphi\right\rangle =\left\langle V_{2}G_{n-1}^{(\theta)}(\cdot,\cdot|y,t_{0}),\varphi\right\rangle \label{Eq4}\end{equation}
 cf.~\cite{LLSW93}, \cite{KS92}. The generic case (\ref{Eq2})
is intermediate between (\ref{Eq3}) and (\ref{Eq4}) and is dealt
with by a combination of the corresponding techniques. This yields
\begin{eqnarray*}
 &  & \left\langle \int_{t_{0}}^{\cdot}d^{n-k}s\int_{\Bbb{R}^{n-k}}\prod_{l=1}^{n-k}dm_{1}(\alpha_{l})G_{k}^{(\theta_{n-k})}(\cdot,\cdot|y,t_{0}),\hat{L}^{*}\varphi\right\rangle \\
 & = & i(n-k)\left\langle V_{1}\int_{t_{0}}^{\cdot}d^{n-k-1}s\int_{\Bbb{R}^{n-k-1}}\prod_{l=1}^{n-k-1}dm_{1}(\alpha_{l})G_{k}^{(\theta_{n-k-1})}(\cdot,\cdot|y,t_{0}),\varphi\right\rangle \\
 &  & +\left\langle V_{2}\int_{t_{0}}^{\cdot}d^{n-k}s\int_{\Bbb{R}^{n-k}}\prod_{l=1}^{n-k}dm_{1}(\alpha_{l})G_{k-1}^{(\theta_{n-k})}(\cdot,\cdot|y,t_{0}),\varphi\right\rangle ,\end{eqnarray*}
 for any $k=2,...,n-2$, \begin{eqnarray*}
 &  & \left\langle \int_{t_{0}}^{\cdot}d^{n-1}s\int_{\Bbb{R}^{n-1}}\prod_{l=1}^{n-1}dm_{1}(\alpha_{l})G_{1}^{(\theta_{n-1})}(\cdot,\cdot|y,t_{0}),\hat{L}^{*}\varphi\right\rangle \\
 & = & i(n-1)\left\langle V_{1}\int_{t_{0}}^{\cdot}d^{n-2}s\int_{\Bbb{R}^{n-2}}\prod_{l=1}^{n-2}dm_{1}(\alpha_{l})G_{1}^{(\theta_{n-2})}(\cdot,\cdot|y,t_{0}),\varphi\right\rangle \\
 &  & +\left\langle V_{2}\int_{t_{0}}^{\cdot}d^{n-1}s\int_{\Bbb{R}^{n-1}}\prod_{l=1}^{n-1}dm_{1}(\alpha_{l})K_{0}^{(\theta_{n-1})}(\cdot,\cdot|y,t_{0}),\varphi\right\rangle ,\end{eqnarray*}
 and \begin{eqnarray*}
 &  & \left\langle \int_{t_{0}}^{\cdot}ds\int_{\Bbb{R}}dm_{1}(\alpha_{1})G_{n-1}^{(\theta_{1})}(\cdot,\cdot|y,t_{0}),\hat{L}^{*}\varphi\right\rangle \\
 & = & i\left\langle V_{1}G_{n-1}^{(\theta)}(\cdot,\cdot|y,t_{0}),\varphi\right\rangle +\left\langle V_{2}\int_{t_{0}}^{\cdot}ds\int_{\Bbb{R}}dm_{1}(\alpha_{1})G_{n-2}^{(\theta_{1})}(\cdot,\cdot|y,t_{0}),\varphi\right\rangle .\end{eqnarray*}
As a result \begin{eqnarray*}
 &  & \left\langle \hat{L}K_{n}^{(\theta)},\varphi\right\rangle \\
 & = & \frac{(-i)^{n-1}}{(n-1)!}\left\langle \left(V_{1}+V_{2}\right)\int_{t_{0}}^{\cdot}d^{n-1}s\int_{\Bbb{R}^{n-1}}\prod_{l=1}^{n-1}dm_{1}(\alpha_{l})K_{0}^{(\theta_{n-1})}(\cdot,\cdot|y,t_{0}),\varphi\right\rangle \\
 &  & +\sum_{k=1}^{n-2}\frac{(-i)^{n-k-1}}{(n-k-1)!}\left\langle V_{1}\int_{t_{0}}^{\cdot}d^{n-k-1}s\int_{\Bbb{R}^{n-k-1}}\prod_{l=1}^{n-k-1}dm_{1}(\alpha_{l})G_{k}^{(\theta_{n-k-1})}(\cdot,\cdot|y,t_{0}),\varphi\right\rangle \\
 &  & +\sum_{k=2}^{n-1}\frac{(-i)^{n-k}}{(n-k)!}\left\langle V_{2}\int_{t_{0}}^{\cdot}d^{n-k}s\int_{\Bbb{R}^{n-k}}\prod_{l=1}^{n-k}dm_{1}(\alpha_{l})G_{k-1}^{(\theta_{n-k})}(\cdot,\cdot|y,t_{0}),\varphi\right\rangle \\
 &  & +\left\langle \left(V_{1}+V_{2}\right)G_{n-1}^{(\theta)}(\cdot,\cdot|y,t_{0}),\varphi\right\rangle ,\end{eqnarray*}
 which is equivalent to \[
\left\langle \hat{L}K_{n}^{(\theta)},\varphi\right\rangle =\left\langle \left(V_{1}+V_{2}\right)K_{n-1}^{(\theta)},\varphi\right\rangle ,\quad\varphi\in\mathcal{D}(\Omega),\]
 for any $n\geq1$. Using (\ref{free}) and summing over $n$, we obtain 
(\ref{Eq5}).\hfill$\blacksquare \medskip$

We conclude by an observation which is obvious from the above construction
but somewhat unexpected given that the Hamiltonians with potentials
in the class $\mathcal{K}_{2}$ will in general not admit a perturbative
expansion (see e.g. \cite{KSW98} for more on this).

\begin{prop}
For any potential $V=g\left(V_{1}+V_{2}\right)$ with
$V_{i}\in\mathcal{K}_{i}$, the solution $K$ of the propagator equation 
(\ref{FOS}) is analytic in the coupling constant $g$.
\end{prop}

\subsection*{Acknowledgments}

M.J.O.~would like to express her gratitude to Jos\'{e} Luís da Silva
for helpful discussions and also the generous hospitality of Cust\'{o}dia
Drumond and CCM during a very pleasant stay at Funchal during the
Madeira Math Encounters XXIII. This work was supported by FCT POCTI, FEDER. 

\section*{Appendix: An estimate}

For the proof of Proposition \ref{Prop1}, we need to estimate\begin{eqnarray*}
 &  & \left|T\Phi_{n,k}(\theta)\right|\\
 & \leq & \exp\!\!\left(\left|x\right|\sum_{l=1}^{n-k}\left|\alpha_{l}\right|\right)\\
 &  & \times\prod_{j=1}^{k+1}\frac{1}{\sqrt{2\pi(\tau_{j}-\tau_{j-1})}}\left|\exp\!\!\left(-\frac{i}{2}\int_{\Bbb{R}}\theta^{2}(s)ds+\sum_{l=1}^{n-k}\alpha_{l}\int_{\Bbb{R}}\theta(s)1\!\!1_{(s_{l},t]}(s)ds\right)\right|\\
 &  & \times\left|\exp\!\!\left(\sum_{j=1}^{k+1}\frac{i}{2\left(\tau_{j}-\tau_{j-1}\right)}\left(\int_{\tau_{j-1}}^{\tau_{j}}\theta(s)ds\right)^{2}\right)\right|\\
 &  & \times\left|\exp\!\!\left(\sum_{j=1}^{k+1}\frac{1}{\tau_{j-1}-\tau_{j}}\left(\int_{\tau_{j-1}}^{\tau_{j}}\theta(s)ds\right)\sum_{l=1}^{n-k}\alpha_{l}\left(\int_{\tau_{j-1}}^{\tau_{j}}1\!\!1_{(s_{l},t]}(s)ds\right)\right)\right|\\
 &  & \times\left|\exp\!\!\left(\sum_{j=1}^{k+1}\frac{i\left(x_{j}-x_{j-1}\right)}{\tau_{j}-\tau_{j-1}}\int_{\tau_{j-1}}^{\tau_{j}}\left(\theta(s)+i\sum_{l=1}^{n-k}\alpha_{l}1\!\!1_{(s_{l},t]}(s)\right)ds\right)\right|\end{eqnarray*}
 We shall now estimate, consecutively, the exponents occuring in the
above expression. 

Using the Cauchy-Schwarz inequality we may approximate\begin{eqnarray*}
 &  & \left|\exp\left(\sum_{l=1}^{n-k}\alpha_{l}\int_{\Bbb{R}}\theta(s)1\!\!1_{(s_{l},t]}(s)ds\right)\right|\\
 & \leq & \exp\left(\sum_{l=1}^{n-k}|\alpha_{l}|\left(\int_{\Bbb{R}}|\theta(s)|^{2}ds\right)^{1/2}\sqrt{t-s_{l}}\right)\\
 & \leq & \exp\left(\sqrt{t-t_{0}}|\theta|\sum_{l=1}^{n-k}|\alpha_{l}|\right)\end{eqnarray*}
 and, similarly, \[
\left|\sum_{j=1}^{k+1}\frac{i}{2\left(\tau_{j}-\tau_{j-1}\right)}\left(\int_{\tau_{j-1}}^{\tau_{j}}\theta(s)ds\right)^{2}\right|\leq\frac{1}{2}|\theta|^{2},\]
as well as\begin{eqnarray*}
 &  & \left|\sum_{j=1}^{k+1}\frac{1}{\tau_{j-1}-\tau_{j}}\left(\int_{\tau_{j-1}}^{\tau_{j}}\theta(s)ds\right)\sum_{l=1}^{n-k}\alpha_{l}\left(\int_{\tau_{j-1}}^{\tau_{j}}1\!\!1_{(s_{l},t]}(s)ds\right)\right|\\
 & \leq & \sum_{j=1}^{k+1}\frac{1}{\tau_{j}-\tau_{j-1}}\left(\int_{\tau_{j-1}}^{\tau_{j}}|\theta(s)|ds\right)(\tau_{j}-\tau_{j-1})\sum_{l=1}^{n-k}|\alpha_{l}|\\
 & = & \sum_{l=1}^{n-k}|\alpha_{l}|\int_{t_{0}}^{t}|\theta(s)|ds\leq\sqrt{t-t_{0}}|\theta|\sum_{l=1}^{n-k}|\alpha_{l}|,\end{eqnarray*}
 where we have again used the Cauchy-Schwarz inequality to obtain
the latter inequality. 

Finally, in order to estimate the exponential of the function \begin{eqnarray*}
 &  & \sum_{j=1}^{k+1}\frac{i\left(x_{j}-x_{j-1}\right)}{\tau_{j}-\tau_{j-1}}\int_{\tau_{j-1}}^{\tau_{j}}\left(\theta(s)+i\sum_{l=1}^{n-k}\alpha_{l}1\!\!1_{(s_{l},t]}(s)\right)ds\\
 & = & \sum_{j=1}^{k+1}\frac{i\left(x_{j}-x_{j-1}\right)}{\tau_{j}-\tau_{j-1}}\int_{\tau_{j-1}}^{\tau_{j}}\theta(s)ds\\
 &  & +\sum_{l=1}^{n-k}\alpha_{l}\sum_{j=1}^{k+1}\frac{x_{j-1}-x_{j}}{\tau_{j}-\tau_{j-1}}\int_{\tau_{j-1}}^{\tau_{j}}1\!\!1_{(s_{l},t]}(s)ds,\end{eqnarray*}
 first we proceed as in \cite{W95}, i.e., \begin{eqnarray*}
\sum_{j=1}^{k+1}\frac{x_{j}-x_{j-1}}{\tau_{j}-\tau_{j-1}}\int_{\tau_{j-1}}^{\tau_{j}}\theta(s)ds & = & \frac{x}{t-\tau_{k}}\int_{\tau_{k}}^{t}\theta(s)ds-\frac{y}{\tau_{1}-t_{0}}\int_{t_{0}}^{\tau_{1}}\theta(s)ds\\
 &  & +\sum_{j=1}^{k}x_{j}\!\left(\!\frac{\int_{\tau_{j-1}}^{\tau_{j}}\theta(s)ds}{\tau_{j}-\tau_{j-1}}-\frac{\int_{\tau_{j}}^{\tau_{j+1}}\theta(s)ds}{\tau_{j+1}-\tau_{j}}\!\right).\end{eqnarray*}
 By the mean value theorem \[
\sum_{j=1}^{k}x_{j}\left(\frac{\int_{\tau_{j-1}}^{\tau_{j}}\theta(s)ds}{\tau_{j}-\tau_{j-1}}-\frac{\int_{\tau_{j}}^{\tau_{j+1}}\theta(s)ds}{\tau_{j+1}-\tau_{j}}\right)=\sum_{j=1}^{k}x_{j}\left(\theta(r_{j})-\theta(r_{j+1})\right),\]
 where $r_{j}\in(\tau_{j-1},\tau_{j})$. Therefore \begin{eqnarray*}
 &  & \left|\sum_{j=1}^{k+1}\frac{i\left(x_{j}-x_{j-1}\right)}{\tau_{j}-\tau_{j-1}}\int_{\tau_{j-1}}^{\tau_{j}}\theta(s)ds\right|\\
 & \leq & \left(|x|+|y|\right)\sup_{\left[t_{0},t\right]}|\theta|+\max_{1\leq j\leq k}|x_{j}|\sum_{j=1}^{k}\left|\int_{r_{j}}^{r_{j+1}}\dot{\theta}(s)ds\right|\\
 & \leq & 2\max_{0\leq j\leq k+1}|x_{j}|\left(\sup_{\left[t_{0},t\right]}|\theta|+\int_{t_{0}}^{t}\left|\dot{\theta}(s)\right|ds\right).\end{eqnarray*}

Now let us consider the sum \[
\sum_{l=1}^{n-k}\alpha_{l}\sum_{j=1}^{k+1}\frac{x_{j-1}-x_{j}}{\tau_{j}-\tau_{j-1}}\int_{\tau_{j-1}}^{\tau_{j}}1\!\!1_{(s_{l},t]}(s)ds.\]
 Since $s_{l}\in\left[t_{0},t\right]$, there is a $j_{0}\in\{0,1,...,k\}$
such that $s_{l}\in\left[\tau_{j_{0}},\tau_{j_{0}+1}\right]$. This
fact allows to rewrite the second sum in the latter expression as
\[
x_{j_{0}+1}-x+(x_{j_{0}+1}-x_{j_{0}})\frac{s_{l}-\tau_{j_{0}+1}}{\tau_{j_{0}+1}-\tau_{j_{0}}}\]
 leading to \[
\left|\sum_{l=1}^{n-k}\alpha_{l}\sum_{j=1}^{k+1}\frac{x_{j-1}-x_{j}}{\tau_{j}-\tau_{j-1}}\int_{\tau_{j-1}}^{\tau_{j}}1\!\!1_{(s_{l},t]}(s)ds\right|\leq4\max_{0\leq j\leq k+1}|x_{j}|\sum_{l=1}^{n-k}|\alpha_{l}|.\]

Inserting these estimates we obtain\begin{eqnarray*}
 &  & \left|T\Phi_{n,k}(\theta)\right|\\
 & \leq & \exp\!\!\left(\left|x\right|\sum_{l=1}^{n-k}\left|\alpha_{l}\right|\right)\prod_{j=1}^{k+1}\frac{1}{\sqrt{2\pi(\tau_{j}-\tau_{j-1})}}\exp\!\left(\left|\theta\right|^{2}\right)\exp\!\!\left(2\sqrt{t-t_{0}}\left|\theta\right|\sum_{l=1}^{n-k}\left|\alpha_{l}\right|\right)\\
 &  & \!\times\!\exp\!\!\left(2\max_{0\leq j\leq k+1}\left|x_{j}\right|\left(\sup_{\left[t_{0},t\right]}\left|\theta\right|+\int_{t_{0}}^{t}\left|\dot{\theta}(s)\right|ds\right)\!\right)\exp\!\!\left(4\max_{0\leq j\leq k+1}\left|x_{j}\right|\sum_{l=1}^{n-k}\left|\alpha_{l}\right|\right).\end{eqnarray*}

Now we introduce the norm \[
\left\Vert \theta\right\Vert :=\sup_{s\in\left[t_{0},t\right]}\left|\theta(s)\right|+\int_{t_{0}}^{t}\left|\dot{\theta}(s)\right|ds+\left|\theta\right|\]

With respect to this norm one may bound the  previous expression by
\begin{eqnarray*}
 &  & \exp\left(\left|x\right|\sum_{l=1}^{n-k}\left|\alpha_{l}\right|\right)\prod_{j=1}^{k+1}\frac{1}{\sqrt{2\pi(\tau_{j}-\tau_{j-1})}}\exp\left(\left\Vert \theta\right\Vert ^{2}\right)\exp\left(2\sqrt{t-t_{0}}\left\Vert \theta\right\Vert \sum_{l=1}^{n-k}\left|\alpha_{l}\right|\right)\\
 &  & \times\exp\left(2\max_{0\leq j\leq k+1}\left|x_{j}\right|\left\Vert \theta\right\Vert \right)\exp\left(4\max_{0\leq j\leq k+1}\left|x_{j}\right|\sum_{l=1}^{n-k}\left|\alpha_{l}\right|\right).\end{eqnarray*}
 Then we use \[
\sqrt{t-t_{0}}\left\Vert \theta\right\Vert \leq\frac{1}{2}(t-t_{0}+\left\Vert \theta\right\Vert ^{2})\]
 and \[
2\max_{0\leq j\leq k+1}\left|x_{j}\right|\left\Vert \theta\right\Vert \leq\max_{0\leq j\leq k+1}\left(\left|x_{j}\right|^{2}\right)+\left\Vert \theta\right\Vert ^{2}\]
to obtain the desired estimate (\ref{po}).

\end{document}